# STANDING WAVES IN THE UNIVERSE

**Evangelos Chaliasos**

365 Thebes Street

GR-12241 Aegaleo

Athens, Greece

*Abstract*

At first, a review of our knowledge on the distribution of galaxies at large-scale, leading to a foam-like large-scale structure of the Universe, is presented in the Introduction. Then, it is shown how, according to the present theory for the formation of superclusters, wave scalar perturbations of the same frequency traveling in opposite directions give rise to standing waves, which cause a motion of the cosmic material towards the nodes, resulting in the concentration of the cosmic material around the nodes. Generalizing this effect to two (three) dimensions, the cosmic material is concentrated around the node lines (node surfaces). It is proposed that the three-dimensional effect is responsible for the foam-like large-scale structure of the Universe.

## 1. Introduction

The first systematic study on the distribution of galaxies at large-scale was done by G.O. Abell in his doctoral dissertation at Caltech, and it was published in *Astrophysical Journal (Supplement Series)* in 1958, under the title “The distribution of Rich Clusters of Galaxies” (Abell, 1958). He found that galaxies are clumped in clusters in general. A fact which was extracted from the data was that the surface distribution of the centers of the clusters was not random, both when the clusters considered belonged to all distances indiscriminately, and when the clusters considered belonged to various concrete distances. The angular scale of the clumping of the distribution varies roughly inversely proportional with distance, so that the linear scale is the same for all distances. Thus Abell´s data suggest the existence of second order clusters, that is clusters consisted of clusters of galaxies, or, as it prevailed to be said, superclusters of galaxies. In fact, a statistical test, done by Abell, revealed that there was not incompatibility between the distribution observed and a distribution in which all clusters belonged to superclusters. Thus, Abell´s work revealed the distribution of galaxies at large-scale. This was done by using *apparent magnitudes* as distance indicators. But apparent magnitudes are less accurate than *redshifts* as distance indicators.

The main work which aimed directly to the investigation of the galaxy distribution at large-scale through redshifts, was done at the Center for Astrophysics (CfA) of the University of Harvard during many years up to 1982, when its results were published in the form of a paper (Davis *et al,* 1982). The authors of this paper wanted to make a complete sky survey at moderate depth and of large angular size. Thus the space studied constituted a sufficient space sample of the Universe. The CfA survey confirmed that galaxies form clusters in general, and it depicted the existence of superclusters of size 20 – 30 $h^{-1}$Mpc in the form of filaments and lumps. The CfA sky survey also depicted the existence of *voids,* that is great regions not containing galaxies. Various maps resulted by the survey are also presented in the paper mentioned. These maps confirmed the general impression that the clumping leads too clusters at small scale. But, at large-scales, they are characterized by formations (superclusters) linked to one another and surrounding huge voids of size frequently 20 Mpc or more. This distribution of galaxies is evident in the maps at largest scale, and is of hierarchical nature. In particular, the great voids, of about 30 Mpc in diameter, are equally well seen as compared to the clusters and superclusters. Thus it is evident that strong clumping at scale of the order of 20 Mpc is the rule, and that voids of 20 – 30 Mpc in diameter are very frequent. The limits of the voids are defined by links of filaments among the clusters, which possess an obviously composite topology. Thus, our picture of the Universe at large-scale has suffered an important evolution. Long ago, the Abell clusters were considered as isolated rare islands of high density in an otherwise uniform background of field galaxies. Now, it is evident that such a field component does not exist and that the clusters have such a great extension that it is impossible for someone to define where one ends and another begins. Almost all galaxies in the sample are concentrated into clusters at small scales, and there are extended zones with galaxy density less than he mean. In conclusion, the distribution of galaxies in space is characterized by great filamentary superclusters of an extend up to 60 Mpc, and corresponding great voids, in a way such that it resembles to foam, or sponge, of a huge scale, the cells of which are the voids, which contain in their walls concentrated all the galactic material, that is the superclusters. Maps produced by n-body simulations are also presented in the paper mentioned. These maps are roughly fitted to the density and the amplitude of the galactic clumping, but they fail to present the foamy, or spongy, nature of the actual clumping. Thus, all these results present a severe challenge to all theories of galaxy and cluster formation.

The galaxy redshift sky survey of CfA was extended. Some of the results of this extension are presented in Lapparent *et al,* 1986. The new sample presents, as a main characteristic of the new data, the fact that galaxies seem to be on the surfaces of structures resembling in form to bubbles. These bubbles have a typical diameter ≈ 25 $h^{-1}$Mpc. Another relevant paper is that of K.-H. Schmidt (1983). Here also the distribution of galaxy clusters reveals many chains and filaments of galaxies 50 – 200 $h^{-1}$Mpc in size. There are also many voids, some of which were also previously known. H.J. Rood´s review under the title "Clusters of Galaxies" (Rood, 1981) is also relevant to the distribution of galaxies at large-scale in such a way that they form clusters, which in turn join themselves into superclusters leaving huge voids among them. It is recorded in the abstract of this paper that redshift surveys for great homogeneous samples of galaxies have led to an improved picture of the three-dimensional distribution of galaxies in space, and that the basic new recognized characteristics of large-scale are superclusters with typical sizes ≈ 100 Mpc (330 million light years) and also equally large voids free of galaxies. It is stated by Rood (1981) that the three-dimensional order of galaxies seems to be chains and lumps separated by huge voids. In fact, the huge voids are a general phenomenon, as the CfA sky survey´s sample also reveals. I also mention J.H. Oort´s review under the title "Superclusters" (Oort, 1983). He comments various general properties of superclusters like the voids formed among them, and he also mentions the existing supercluster formation theories. It must be observed that, according to these theories, we can obtain the foam-like large-scale structure of the Universe, by performing suitable simulations. For example see Fig.17 in Rood, 1981, and Fig.6 in Einasto *et al,* 1980, reproduced here as Fig.1 and Fig.2 respectively. I will propose in the next section a new simple and natural theory for mainly the formation of the foam-like large-scale structure of the Universe based on the relativistic theory of perturbations (gravitational instability) in the Universe, this latter initiated by Lifshitz & Khalatnikov, 1963.

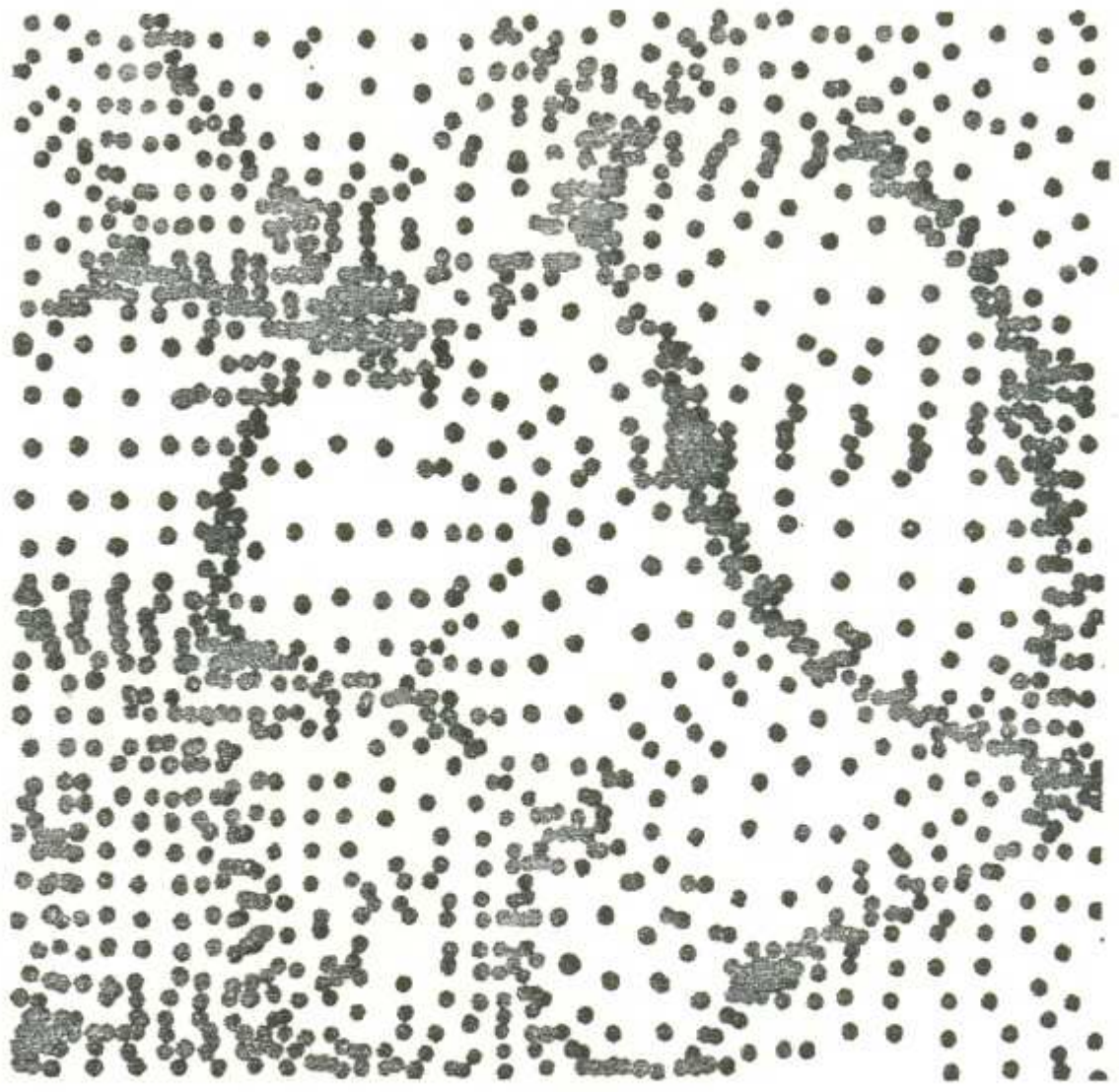

**Figure 17.** By considering the effects of small density perturbations in the early Universe, Zeldovich produced this theoretical model of the distribution of galaxies in the present Universe (from Zeldovich 1978).

Fig.1

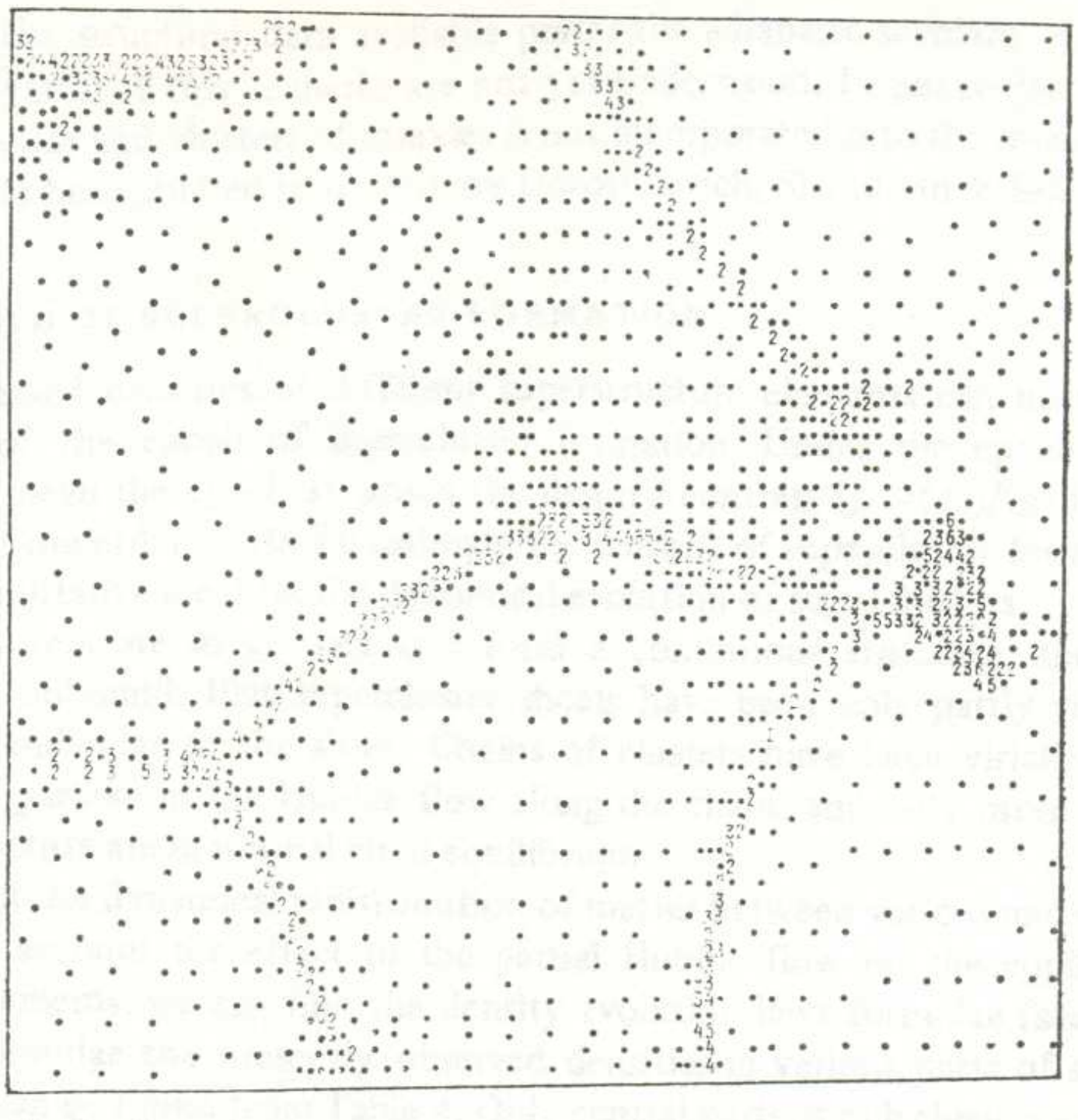

**Figure 6.** Typical distribution of test particles at the end of the supercluster formation according to a two-dimensional numerical experiment (Doroshkevich *et al.* 1976). The numbers indicate the number of test particles in respective places. The concentration of test particles into caustic configurations is clearly seen.

Fig.2

## 2. The theory of standing waves in the Universe

The theory of gravitational instability of an isotropic universe was first developed by Lifshitz & Khalatnikof, 1963. A simplified version of this theory can be found in Landau & Lifshitz, 1975, in Zel´dovich & Novikov, 1983, and in Peebles, 1980. According to this theory the cosmological perturbations can be expanded in plane waves. Using cartesian coordinates we can write the periodic space factor for these waves in the form

$$Q \equiv e^{i\vec{\kappa}\cdot\vec{x}}, \tag{0}$$

where

$$\vec{x} \equiv \vec{r} \equiv \vec{l}/a, \tag{1}$$

and

$$\vec{\kappa} \equiv \vec{n} \equiv a\vec{k}, \tag{2}$$

with a the scale factor, **l** the space distance, and **k** the wave vector. Thus, we obtain

$$Q = e^{i\vec{k}\cdot\vec{l}}. \tag{3}$$

Gravitational perturbations are divided into three types. This classification is as follows.

1. Scalar perturbations, which result in perturbations δε of the energy density, in perturbations of the spatial part $u^\alpha$ of the velocity four-vector (the unperturbed four-velocity is zero for the coordinates are supposed to be comoving), and in perturbations $h_{ik}$ of the metric tensor $g_{ik}$ (that is gravitational waves).[(*)]
2. Vector (or Rotational) perturbations, which result in perturbations $u^\alpha$ of the spatial part of the velocity four-vector, and in perturbations $h_{ik}$ of the metric

---

[(*)] Latin indices (i, k, …) refer to space-time taking the values 0, 1, 2, 3, while Greek indices (α, β, …) refer to space only taking the values 1, 2, 3.

tensor $g_{ik}$ (that is gravitational waves). There are no density perturbations in this type. And

3. Tensor perturbations, which contain only perturbations $h_{ik}$ of the metric tensor $g_{ik}$ (gravitational waves).

We are interested in scalar perturbations. We define the vector

$$\vec{P} \equiv \vec{\kappa}\, Q/\kappa\,, \tag{4}$$

and the tensors

$$Q_{\alpha}{}^{\beta} \equiv \frac{1}{3}\delta_{\alpha}{}^{\beta} Q \tag{5}$$

and

$$P_{\alpha}{}^{\beta} \equiv \left(\frac{1}{3}\delta_{\alpha}{}^{\beta} - \frac{\kappa_{\alpha}\kappa^{\beta}}{\kappa^{2}}\right)Q. \tag{6}$$

For the case that

$$1/\kappa << \eta << 1, \tag{7}$$

where η is the time parameter, given by

$$a d\eta = c dt, \tag{8}$$

where t is the world time (and c is the velocity of light), and for the equation of state (appropriate for this case)

$$p = \varepsilon/3, \tag{9}$$

with p the pressure, we find for the perturbations

$$h_{\alpha}{}^{\beta} = \frac{C}{\kappa^{2}\eta^{2}}\left(P_{\alpha}{}^{\beta} - 2Q_{\alpha}{}^{\beta}\right)e^{i\kappa\eta/\sqrt{3}}, \tag{10}$$

$$\delta\varepsilon/\varepsilon = -(C/9)Qe^{i\kappa\eta/\sqrt{3}}, \tag{11}$$

and

$$u^{\alpha}/c = (C/12\sqrt{3})P^{\alpha}e^{i\kappa\eta/\sqrt{3}}, \tag{12}$$

where C is a linear function of κ, namely

$$C = (4\sqrt{3})C_2 + (3i\sqrt{3})\kappa\, C_1, \tag{13}$$

with $C_1$ and $C_2$ two constants. Eqns. (10), (11), and (12) refer to the early stages of expansion ($\eta \ll 1$), when the equation of state was given by (9), and for the case that $\kappa\eta \gg 1$. Because of eqn. (12), this wave solution of the perturbation equations is about a longitudinal sound wave traveling with a velocity

$$u \equiv \sqrt{\frac{dp}{d(\varepsilon / c^2)}} = c/\sqrt{3}. \tag{14}$$

In fact, if we insert eqn. (14) into eqn. (12), we find for the exponent of the periodic time factor

$$\kappa\eta/\sqrt{3} = \omega\tau, \tag{15}$$

where $\omega$ is the cyclic frequency and $\tau$ is given by

$$\tau \equiv a\int \frac{dt}{a}. \tag{16}$$

We thus have from eqn. (12)

$$\vec{u}/c = (C / 12\sqrt{3})\frac{\vec{\kappa}}{\kappa} Q e^{i\omega\tau}, \tag{17}$$

or, because of eqn. (3),

$$\vec{u} \propto \vec{\kappa}\, e^{i(\omega\tau + \vec{k}\cdot\vec{l}\,)}, \tag{18}$$

that is (taking the real part), in one dimension,

$$u \propto \cos(\omega\tau + kl). \tag{19}$$

Limiting ourselves now to the non-relativistic approximation we can write

$$u \equiv \frac{v}{\sqrt{1 - v^2/c^2}} \cong v, \tag{20}$$

where v is the usual newtonian velocity. Thus we are finally left with the formula

$$v \propto \cos(\omega\tau + kl), \tag{21}$$

or

$$v_1 = K\cos(\omega\tau + kl), \tag{22}$$

with K a function of k (& $v_1 = v$). This is a wave moving in the *negative* **l** direction.

Now, because of the *isotropy of space*, we can always assume that there is also another wave, of the same wave length (and cyclic frequency) and amplitude, moving in the *positive* **l** direction (*opposite* to the wave given by eqn. (22)), namely

$$v_2 = K\cos(\omega\tau - kl). \tag{23}$$

But, because of the *principle of superposition*, the wave (23) *interferes* with the wave (22), so that we will have for the total velocity

$$v_{o\lambda} = v_1 + v_2. \tag{24}$$

Adding the trigonometric functions relevant to $v_1$ and $v_2$ (the cosines), we are left with

$$v_{o\lambda} = 2K\cos(kl)\cos(\omega\tau). \tag{25}$$

This represents a *standing wave* (see Fig.19-17 of Resnick & Halliday, 1966, reproduced here as Fig.3). Each oscillating material point of unit mass (say) will

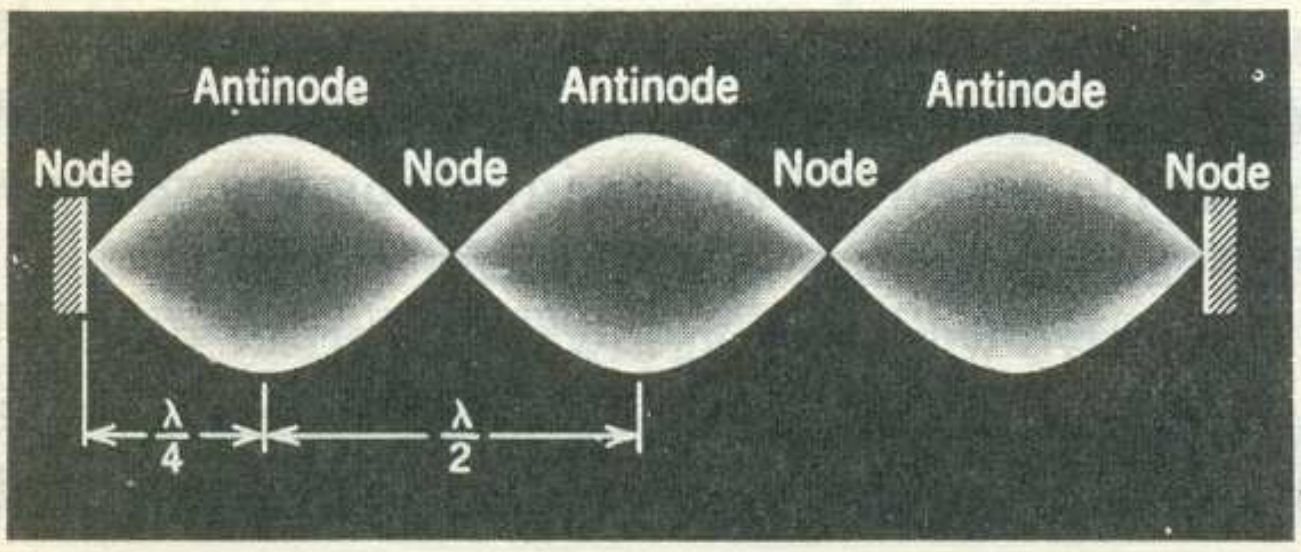

**Fig. 19–17** The envelope of a standing wave, corresponding to a time exposure of the motion, and showing the patterns of nodes and antinodes.

Fig.3

possess a total energy

$$E_{o\lambda} = (1/2)(2K\cos kl)^2, \tag{26}$$

or

$$E_{o\lambda} = 2K^2\cos^2(kl). \tag{27}$$

This total energy depends on the location (l). Thus we can in turn consider it as the *potential energy* of a motion (*other* than the oscillation) along the **l** axis, that is we set

$$U(l) = 2K^2\cos^2(kl). \tag{28}$$

Thus a *force* is acting on the material point (*other* than the restoring force) giving it an acceleration

$$\gamma(l) = -\partial U/\partial l\,. \tag{29}$$

Performing the elementary differentiation in eqn. (29), we obtain

$$\gamma(l) = 4kK^2 \cos(kl)\sin(kl), \tag{30}$$

or

$$\gamma(l) = 2kK^2 \sin(2kl). \tag{31}$$

This acceleration leads the material point under consideration towards the nearest *node* (see Fig.3). Thus the material point will oscillate (but <u>not</u> harmonically!) about a node. Finally, because of *friction,* the material point will settle down at the node. In this way all of the oscillating (<u>harmonically</u> now!) material will be concentrated at the nodes. Thus a simple "structure" is formed at a node, for example a cluster of galaxies.

Up to now we have limited ourselves to *one* dimension and obtained the formation of *point* concentrations of cosmic material, that is "point" structures (clusters). A similar analysis can be performed in *two* dimensions, leading now to concentration of cosmic material on the node *lines* (see Fig.20-7 of Resnick & Halliday, 1966, reproduced here as Fig.4, and Figs.362 & 363 of Alexopoulos, 1960, reproduced here as Figs.5 & 6). Thus "line" structures can be formed, for example filaments.

Finally performing the above analysis in full *three* dimensions, we will obtain a concentration of the cosmic material on the node *surfaces* now. In this way we can simply and naturally explain the existence of huge voids in the Universe, with the cosmic material concentrated on the "walls" of the voids. This is write the foam-like structure of the Universe observed.

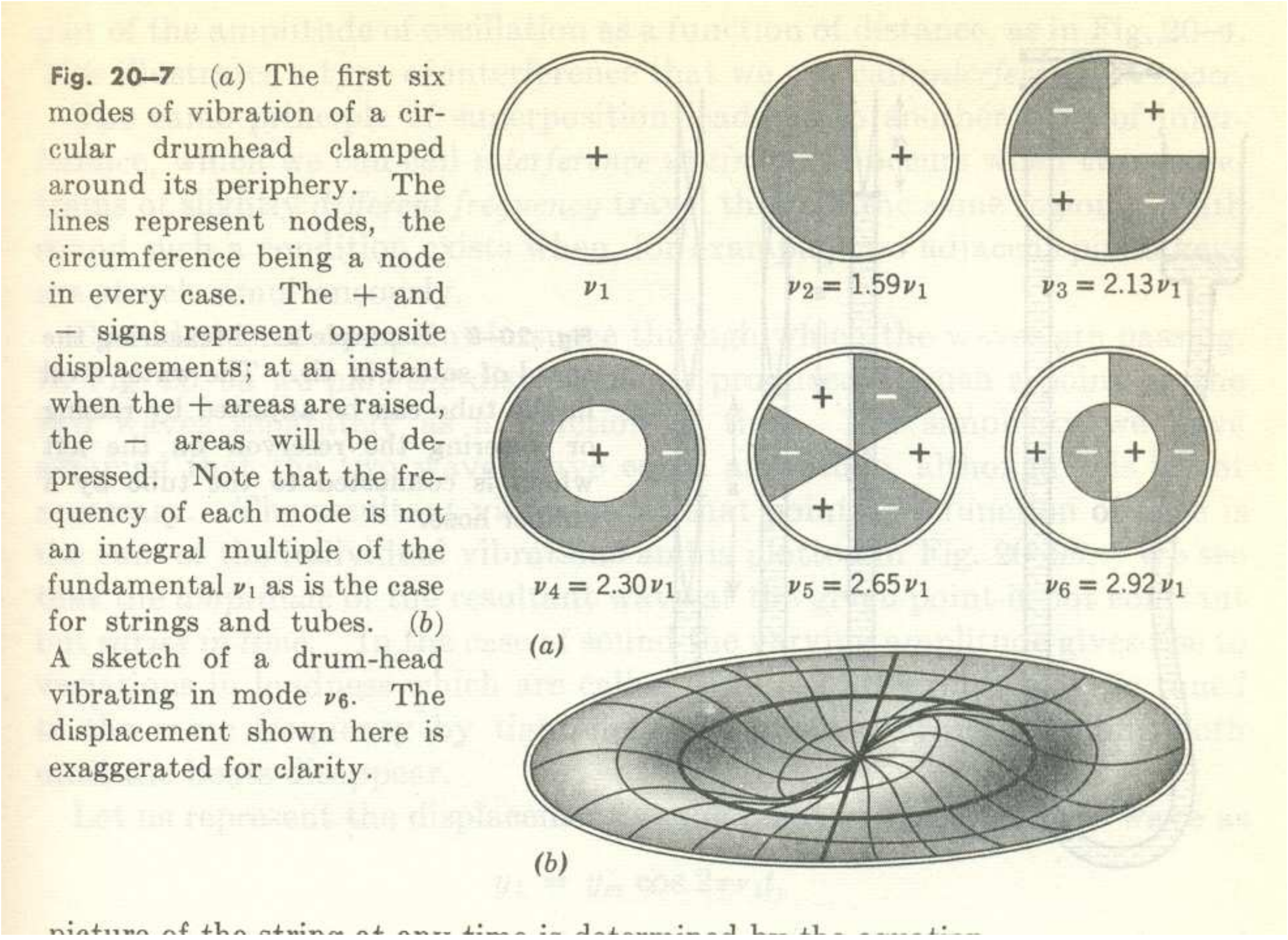

**Fig. 20–7** (*a*) The first six modes of vibration of a circular drumhead clamped around its periphery. The lines represent nodes, the circumference being a node in every case. The + and − signs represent opposite displacements; at an instant when the + areas are raised, the − areas will be depressed. Note that the frequency of each mode is not an integral multiple of the fundamental $\nu_1$ as is the case for strings and tubes. (*b*) A sketch of a drum-head vibrating in mode $\nu_6$. The displacement shown here is exaggerated for clarity.

Fig.4

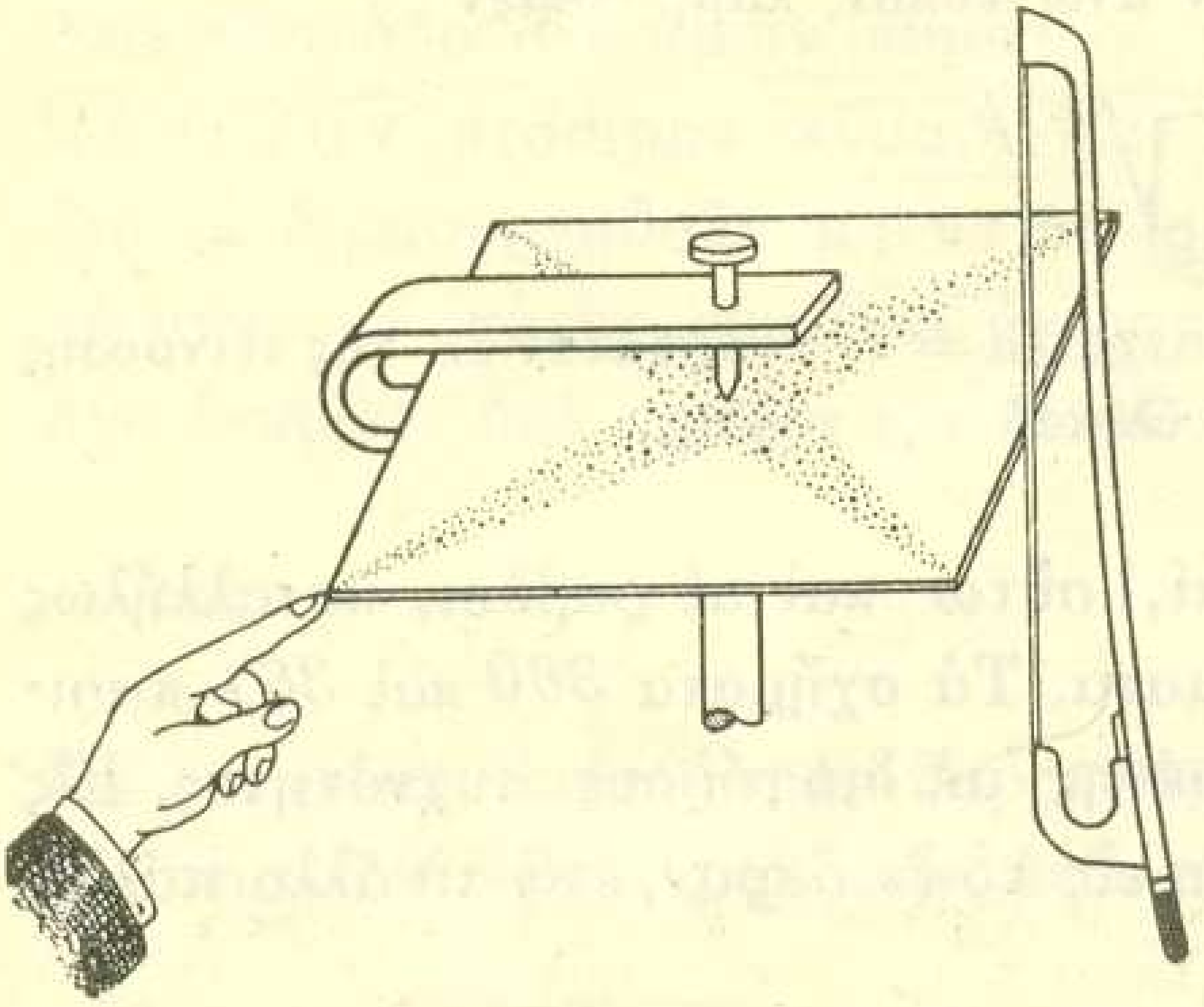

**Σχ. 362.** Ὅταν διὰ τοῦ τόξου διεγείρωμεν τὴν πλάκα, ἡ ἄμμος συγκεντροῦται εἰς τὰς δεσμικὰς γραμμάς. Μία ἐξ αὐτῶν διέρχεται διὰ τοῦ σημείου, τὸ ὁποῖον ἀκινητοποιοῦμεν διὰ τοῦ δακτύλου.

Fig.5

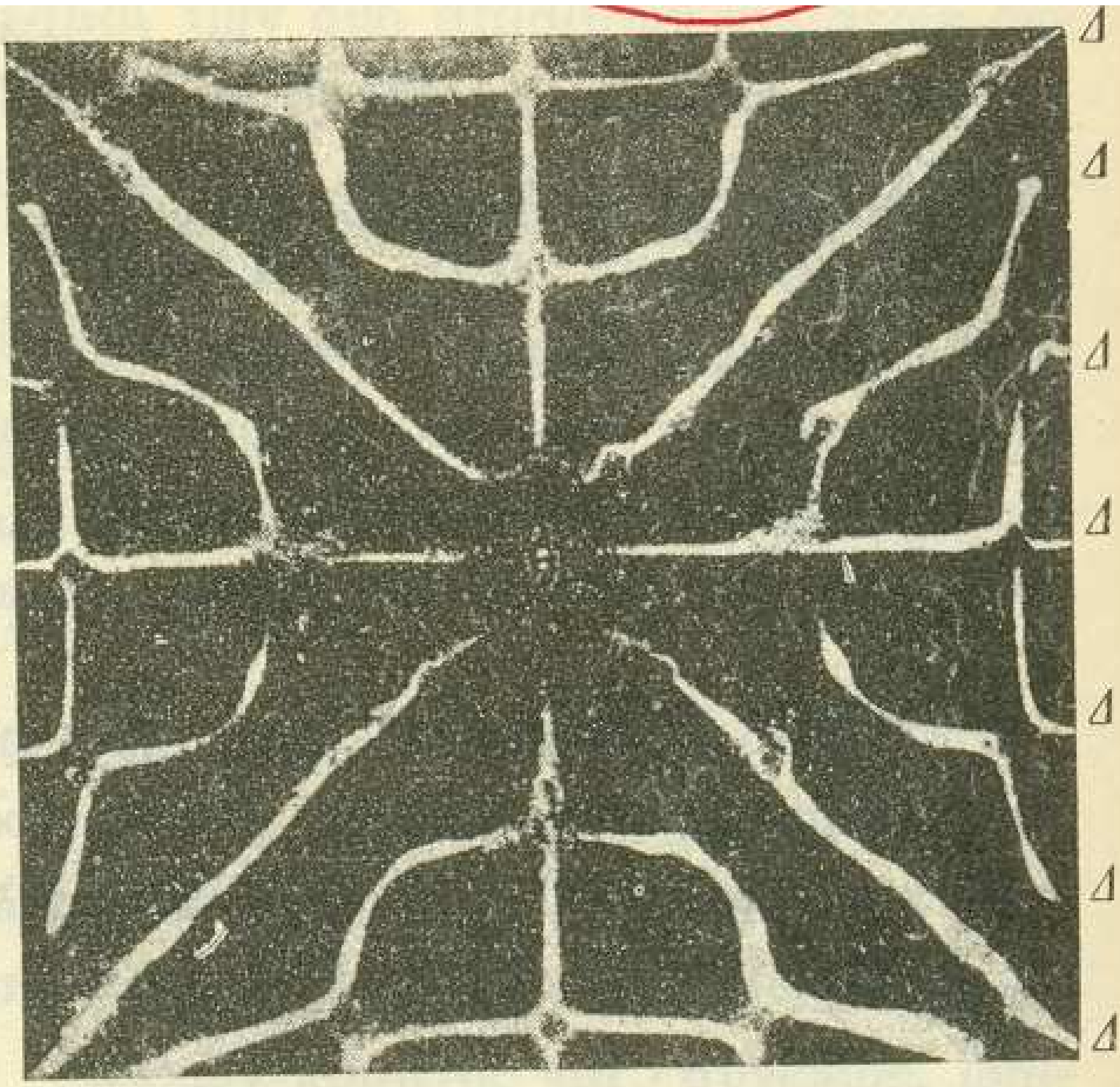

**Σχ. 363.** *Εἰκὼν Chladni, ληφθεῖσα διὰ τῆς μεθόδου, τὴν ὁποίαν παριστᾷ τὸ σχῆμα 362.*
*(Δ, Δ . . . = δεσμικαὶ γραμμαί).*

Fig.6